\documentclass[letterpaper]{article} 
\usepackage{aaai24}  
\usepackage{times}  
\usepackage{helvet}  
\usepackage{courier}  
\usepackage[hyphens]{url}  
\usepackage{graphicx} 
\usepackage{natbib}  
\usepackage{caption} 
\usepackage{algorithm}
\usepackage{algorithmic}

\usepackage{booktabs}
\usepackage{amssymb}
\usepackage{pifont}

\usepackage{amsmath} 
\usepackage{fixltx2e}
\usepackage{paralist}
\usepackage{xcolor}
\usepackage{booktabs}
\usepackage{array}
\usepackage{multirow}
\usepackage{rotating}
\usepackage{arydshln}
\usepackage{amssymb}
\usepackage{pifont}

\usepackage{xcolor}

\usepackage{subcaption}
\usepackage{newfloat}
\usepackage{listings}
\DeclareCaptionStyle{ruled}{labelfont=normalfont,labelsep=colon,strut=off} 
\floatstyle{ruled}
\newfloat{listing}{tb}{lst}{}
\floatname{listing}{Listing}
\title{Hate Cannot Drive out Hate: \\ Forecasting Conversation Incivility following  Replies to Hate Speech}
\author{
	Xinchen Yu\textsuperscript{\rm 1}, 
	Eduardo Blanco\textsuperscript{\rm 1}, 
	Lingzi Hong\textsuperscript{\rm 2}
}
\affiliations{
	\textsuperscript{\rm 1}Department of Computer Science. University of Arizona\\	
	\textsuperscript{\rm 2}Department of Information Science. University of North Texas\\
}

\usepackage{bibentry}

\begin{document}
	
	\maketitle
	
	\begin{abstract}
		User-generated replies to hate speech are promising means to combat hatred,
		but questions about whether they can stop incivility in follow-up conversations linger.
		We argue that effective replies stop incivility from emerging in follow-up conversations---replies that elicit more incivility are counterproductive.
		This study introduces the task of predicting the incivility of conversations following replies to hate speech.
		We first propose a metric to measure conversation incivility based on the number of civil and uncivil comments as well as the unique authors involved in the discourse.
		Our metric approximates human judgments more accurately than previous metrics.
		We then use the metric to evaluate the outcomes of replies to hate speech.
		A linguistic analysis uncovers the differences in the language of replies that elicit follow-up conversations with high and low incivility.
		Experimental results show that forecasting incivility is challenging.
		We close with a qualitative analysis shedding light into the most common errors made by the best model.
	\end{abstract}

	\section{Introduction}
	The pervasive problem of online hate speech has motivated
	researchers to investigate methods for mitigating hatred (e.g., content moderation)~\cite{schmidt-wiegand-2017-survey,10.1145/3232676}.
	Engaging with hate speech---for example by using counterspeech---has recently emerged as an alternative to address hate speech~\cite{richards2000counterspeech}. 
	While content moderation consists in flagging or removing hate speech,
	engaging with hateful content---for example by showing sympathy towards the victim or providing counterarguments---does not
	interfere with the principle of free and open public spaces for debate~\cite{DBLP:conf/icwsm/MathewSTRSMG019,schieb2016governing, chung-etal-2019-conan}.

	The United Nations has suggested reacting to hate speech as an action to deal with hatred.\footnote{https://www.un.org/en/hate-speech/take-action/engage}
	Social media platforms like Facebook have started counterspeech programs.\footnote{\url{https://counterspeech.fb.com/en/}} 
	Recently, the NLP community has focused on analyzing, modeling, and generating replies to hate speech~\cite{DBLP:conf/icwsm/MathewSTRSMG019,tekiroglu-etal-2020-generating,fanton-etal-2021-human,yu-etal-2022-hate}.
	These previous efforts make the following assumption:
	responding to hate speech is an ideal solution to stop or at least mitigate online incivility.
	In this work, incivility is defined as a manner of offensive interaction ranging from abusive language, vulgarity, racism, or sexism, to harassment, name-calling or personal attacks~\cite{antoci2016civility,sadeque-etal-2019-incivility,davidson-etal-2020-developing}.
	While intuitive, we are not aware of strong evidence supporting this assumption.
	Consider the Reddit conversation in Figure \ref{f:problem-example}.\footnote{The examples in this paper contain hateful content. We cannot avoid it due to the nature of our work.}
	The first post is hateful towards feminists in general. 
	The second post (i.e., the reply) appears to be a strong reply that counters the hateful content.
	As strong as it might be, however, the second post elicits additional incivility:
	the third post escalates the hatred further by attacking the author of the the second post.
	
	\begin{figure}[t]
		\small
		\centering
		\begin{tabular}{@{}p{\columnwidth}@{}}
			\toprule
			Hateful post:
			\emph{Just curious how you can identify with a movement which has essentially become a hate group full of crazy feminists.}  \\ \addlinespace
			Reply to hateful post:
			\emph{Come on man, most feminists are ok. Hate group? how can you use such a strong term?}  \\ \addlinespace
			Uncivil post (after the reply):
			\emph{No, it’s not strong. Don’t lie through your teeth, c**t.} \\ 
			\bottomrule
		\end{tabular}
		\caption{
			An excerpt from a Reddit conversation.
			The second post contains counterspeech but it elicits additional uncivil behaviors.
			Indeed, the third post escalates the hate with respect to the original hateful post.
		}
		\label{f:problem-example}
	\end{figure}
	
	At face value, coming up with elaborate counterspeech replies does address online hatred.
	Some replies, however, may elicit additional incivility in the subsequent conversations. 
	Existing work lacks a deeper understanding of what replies to hate speech can stop the spread of hatred and prevent uncivil content from emerging in subsequent conversations.
	In this paper, we aim to computationally assess and forecast conversational outcomes (namely, \emph{conversation incivility}) of replies to hate speech.
	We look at all replies with varying conversational outcomes. 
	Regardless of the content of replies---short or long, offensive or polite, well-argued or fatally flawed from a logical standpoint---we consider the outcome is civil if the discourse that follows is primarily not uncivil.
	Further, we argue that looking at genuine online discourse and assessing what comments elicit additional uncivil behaviors---even if they are well-meaning and polished arguments---is a worthwhile goal.
	
	We focus on incivility as the conversational outcome and investigate how language usage is tied to the future trajectory of a conversation. 
	Recent studies on measuring conversation incivility use number or ratio of uncivil comments~\cite{DBLP:conf/icwsm/LiuGHC18,10.1145/3366423.3380273,DBLP:conf/kdd/DahiyaSSGCEMB021,garland2022impact}.
	In contrast, we quantify conversation incivility by also considering conversation length~\cite{tsagkias2009predicting,Yano_Smith_2010,artzi-etal-2012-predicting} and user re-entry behaviors~\cite{10.1145/2433396.2433401}. 
	This allows us to build bridges between works on conversation modeling in general and in incivility domains.
	Additionally, the work presented here could complement current studies on counterspeech by 
	investigating a richer source of replies to hate speech
	and 
	providing insights into strategies to respond to hatred.
	
	We propose a new metric to assess conversation incivility and apply it to evaluate conversations following replies to hate speech in a new dataset of Reddit conversations.\footnote{Data and code available at anonymous.link}
	Our metric takes into account the number of uncivil and civil comments in a conversation and also considers user behavior.
	Specifically, our metric differentiates between comments by many different ``drive-by" people
	and those by relatively few people who engage multiple times~\cite{10.1145/2433396.2433401}. 
	Intuitively, a reply to hateful content is more divisive and problematic if it attracts more hate speakers to participate, misbehave, or name-call.
	Our metric differentiates replies depending on the outcomes of the follow-up conversation.
	Additionally, as we shall see, the metric approximates human judgments more accurately than previous metrics.
	For instance, the conversation following a reply is less civil if it consists of five uncivil posts instead of one, 
	despite the ratio of uncivil posts is the same. 
	Similarly, if the five uncivil posts are from different users, 
	the conversation is less civil than if they were from the same user, 
	despite the number and ratio of uncivil posts are the same.
	
	We focus on conversation incivility following replies to hate speech in Reddit. 
	A linguistic analysis is presented to show the differences in the language of replies to hate speech
	depending on the incivility of the follow-up conversation.
	Importantly,
	we show that the linguistic insights hold across several Reddit communities,
	including some that are known for respectful debate rather than hateful content (e.g., \textit{r/ChangeMyView}).
	We then experiment with classifiers to predict whether a reply will be followed by a conversation with high, medium or low incivility.
	Our models obtain modest results,
	and we present a qualitative error analysis.
	
	In summary, we answer the following research questions:
	\begin{compactenum}
		\item Do all replies to hateful posts elicit conversations with the same incivility? (they don't);
		\item Do replies that elicit high, medium, and low conversation incivility use different language? (they do);
		\item Do models to predict the incivility of the conversation following replies to hateful post benefit from having access to the hateful post in addition to the reply? (they do);
		\item When differentiating between replies eliciting the top-$k$ highest and lowest incivility,
		is it true that the smaller the~$k$ the easier the task? (it is).
	\end{compactenum}

	\section{Related Work}
	\label{s:related_work}

	\paragraph{Replying to hate speech}
	There is an abundance of research working on replies to hate speech.
	Most prior work focuses on counterspeech and contribute several corpora with counterspeech content~\cite{DBLP:conf/icwsm/MathewSTRSMG019,qian-etal-2019-benchmark,chung-etal-2019-conan,yu-etal-2022-hate}.
	~\citet{chung-etal-2019-conan} collect synthetic counterspeech generated on-demand by trained operators.
	Compared to genuine counterspeech written by regular people out of their own desires and motivations, synthetic counterspeech is more generic and poorly addresses specific hateful content (e.g., \emph{This kind of language is inappropriate and should be avoided}).
	Some user studies that investigate the outcome of replies to hatred lack in scale; 
	they compare the control group with the treatment group that has received interventions~\cite{munger2017tweetment,hangartner2021empathy,bilewicz2021artificial,wachs2023effects}. 
	The only two large-scale previous works are by \citet{garland2022impact} and \citet{albanyan-etal-2023-counterhate}.
	\citet{garland2022impact} calculate the hate score of each reply and estimate its impact by comparing the 
	average hate scores before and after.
	\citet{albanyan-etal-2023-counterhate} focus on the direct replies after each reply to hatred.
	While we are inspired by previous approaches addressing online hatred, we
	analyze \emph{all} user-generated comments after replies to hate speech \emph{in genuine online conversations}.
	Additionally,
	measuring incivility automatically allows us to bypass the burden of manual annotations
	and work with orders of magnitude more data than previous work.
	
	\paragraph{Conversational forecasting}
	There are several efforts on forecasting whether online content will result in additional uncivil behaviors. 
	\citet{10.1145/2998181.2998213} predict whether a moderator will flag a post for removal.
	\citet{zhang-etal-2018-conversations} and \citet{Yuan_Singh_2023} predict whether a few utterances at the very beginning of a conversation will lead to a personal attack.
	\citet{DBLP:conf/icwsm/LiuGHC18} forecast whether an Instagram post will receive more than $n$ uncivil comments.
	\citet{DBLP:conf/kdd/DahiyaSSGCEMB021} forecast the incivility score of upcoming tweet replies.
	Similarly, our work aims to forecast incivility of the conversation following a reply to hate speech. 
	We, however, measure incivility by also considering the total number of comments as well as user re-entry behaviors in a conversation.
	As we shall see, our metric to measure conversation incivility 
	better approximates human judgments.
	
	Besides conversation incivility, a rich source of other conversational outcomes have been explored, including betrayal in games~\cite{niculae-etal-2015-linguistic}, success in persuading others~\cite{10.1145/2872427.2883081}, debate winners~\cite{potash-rumshisky-2017-towards}, online conflicts~\cite{levy2022understanding}, and prosocial behaviors~\cite{10.1145/3442381.3450122,lambert2022conversational}.
	We build on prior work in modeling conversation trajectory from the structural aspects of conversations~\cite{10.1145/2433396.2433401} and complement them by considering linguistic aspects.

	\section{Measuring Conversation Incivility} 
	\label{s:metric}
	We propose a new metric to measure conversation incivility following a reply~$r$.
	Our metric consists of two main components: uncivil behavior $U(r)$ and civil behavior $C(r)$. A comment is either civil or uncivil. 
	All comments in the subsequent conversation after a reply $r$ are from a population of unique users $P$.
	For each user~$i$ in $P$ (for $i = 0,1,2,...,k$), let $n_{ui}$ denote the number of uncivil comments the user $P_i$ posts, and $n_{ci}$ denote the number of civil comments.
	Uncivil behavior $U(r)$ is defined as the sum of $f(n_{ui})$ over all the users $i$ in the conversation after~$r$.
	Similarly, civil behavior $C(r)$ is defined as the sum $f(n_{ci})$ over all the users.
	Here,~$f$ is a strictly increasing and concave down function passing through the origin.
	When there are no comments after a reply $r$, the conversation incivility is equal to 0. 
	Intuitively, the conversation incivility following a reply~$r$ is higher when there are
	(a) more uncivil comments by many people and
	(b) fewer civil comments by a handful of people.
	
	We formally define the conversation incivility score of $r$ ($S(r)$) as follows:
	$$S(r) = \alpha U(r) - (1-\alpha) C(r)$$
	where
	$U(r) =	\sum_{i=1}^{k} f({n_{ui}})$ and
	$C(r) =	\sum_{i=1}^{k} f({n_{ci}})$.
	
	The parameter $\alpha$ determines how much weight to give to each component.
	The larger $\alpha$ is, the more weight is given to uncivil behavior
	(i.e., more civil comments are needed to neutralize one uncivil comment).
	Following the literature, the future trajectory of a conversation becomes more toxic and unhealthy if it involves a larger number of participants venting or misbehaving, and this should be treated differently from situations in which there are only few people with repeated engagement~\cite{10.1145/2433396.2433401}.
	Finally, a reply eliciting a long and civil conversation reflects that it attracts a great deal of attention and promotes healthy discussions. 
	Our metric builds on prior work on comment-volume prediction~\cite{artzi-etal-2012-predicting,10.1145/2433396.2433401}, but we adapt it to model conversation incivility.
	
	In this paper, we experiment with the square root function as $f$
	(i.e., $f(x) = \sqrt{x}$), but any strictly increasing and concave down function is a valid choice.
	The rate of increase of these functions slows as $x$ grows;
	different $f$ choices dictate how quickly additional comments by the same author are deemed unimportant.
	As shown in the appendices,
	the Spearman's correlation coefficients between $S(r)$ using $\sqrt{x}$ and other  choices
	($\log(x)$, $\sqrt[3]{x}$, $arctan(x)$, $tanh(x)$)
	are over 0.95 ($p<0.001$).
	In other words, the choice of $f$ is not critical.
	Indeed, while the absolute scores of conversation incivility will vary depending on $f$,
	comparing scores would lead to the same conclusions.

	\section{A Corpus of Online Hate, Replies, and Follow-Up Conversation Incivility}
	\label{s:corpus}
	
	\begin{figure}[t]
		\centering
		\includegraphics[width=1\linewidth]{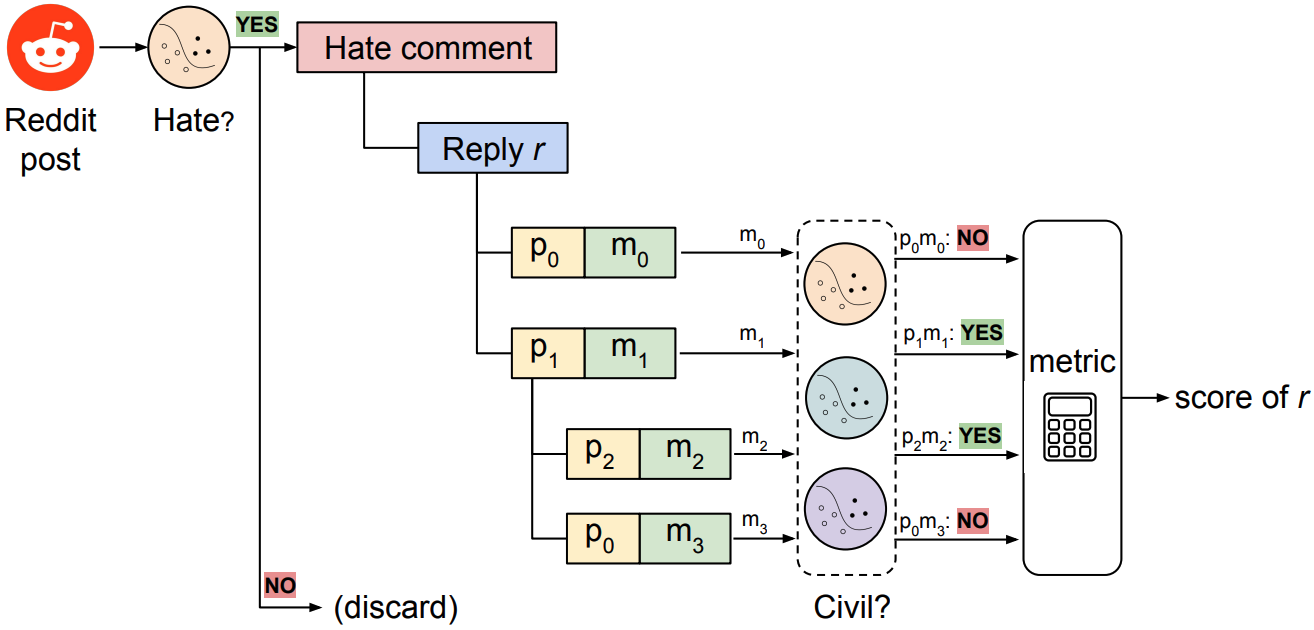}
		\caption{The pipeline to create our corpus. 
			We show a Reddit post, the reply to it, and the subsequent conversation.
			The output is the incivility score of the conversation following the reply.
			We indicate each comment in the subsequent conversation as $m_{j}$ (for $j = 0,1,..,h$) and the user who posts it as $p_{i}$ (for $i = 0,1,..,k$). 
			The pipeline includes two questions: (1) is the Reddit post hateful? and (2) is $m_{j}$ civil?
			We adopt pretrained classifiers to answer the questions. } 
		\label{fig:illustration}
	\end{figure}

	We choose Reddit as the starting point for our corpus 
	and
	use the PushShift API to retrieve whole conversation threads.\footnote{\url{https://pushshift.io/api-parameters/}}
	As the prevalence of online hate in the wild is very low (0.1\% in English language social media~\cite{vidgen-etal-2019-challenges}), many studies use keyword sampling to increase the chances. 
	Keywords, however, may introduce topic and author biases~\cite{wiegand-etal-2019-detection,vidgen-etal-2021-introducing}. 
	Instead, we use community-based sampling and identify
	39 subreddits that are thought to be more or less hateful~\cite{qian-etal-2019-benchmark,guest-etal-2021-expert,vidgen-etal-2021-introducing} or previously used in conversational forecasting~\cite{chang-danescu-niculescu-mizil-2019-trouble,Yuan_Singh_2023} (see Appendices for the full list). 
	Note that some subreddits we work with primarily consist of respectful conversations rather than hateful content (e.g., \textit{r/technology}, \textit{r/changemyview}).
	We retrieve 1,382,596 comments from 5,325 submissions in the 39 subreddits.
	The publication time of the Reddit conversations range from October 16, 2020, to February 20, 2022.
	
	We focus on replies that directly reply to hate speech. 
	Therefore, the next steps are to 
	(a)~identify the comments that are hateful and their replies,
	and
	(b)~assess the incivility of the follow-up conversations.
	The second step requires identifying uncivil content in the comments following the reply.
	Figure~\ref{fig:illustration} illustrates the process.

	\paragraph{Identifying Hate Comments and Their Replies}
	\label{ss:identify}
	We identify hate speech in the 1,382,596 comments using pre-trained models \cite{DBLP:journals/corr/abs-1907-11692} fine-tuned with the corpus by~\citet{qian-etal-2019-benchmark}
	and
	the implementation by~\citet{phang2020jiant}.
	We make this choice for several reasons.
	First, the corpus annotates Reddit comments as hateful or not hateful, the same domain we work with.
	Second, the classifier obtains outstanding results: 0.93 F1.
	In a more strict evaluation using Cohen's $\kappa$, we obtain $\kappa=0.83$ between the predictions and the gold annotations in the test set.
	Note that $\kappa$ coefficients above $0.80$ indicate (almost) perfect agreement~\cite{artstein2008inter}.
	Thus the predictions are reliable enough to be considered ground truth.
	
	After automatically identifying hateful comments, we pair
	(a) each hateful comment with each of its direct replies
	and
	(b) each reply to a hateful comment with the follow-up conversation (i.e., all the subsequent comments in the same thread).
	We found 21,845  hate comments in the 39 subreddits we work with.
	On average, a hate comment has 1.56 direct replies,
	and there are 2.36 comments in the conversation following up a reply to hateful content.

	\paragraph{Assessing Conversation Incivility}
	\label{ss:assess}
	\begin{figure}[t]
		\centering
		\begin{subfigure}[b]{0.23\textwidth}
			\centering
			\includegraphics[width=\textwidth]{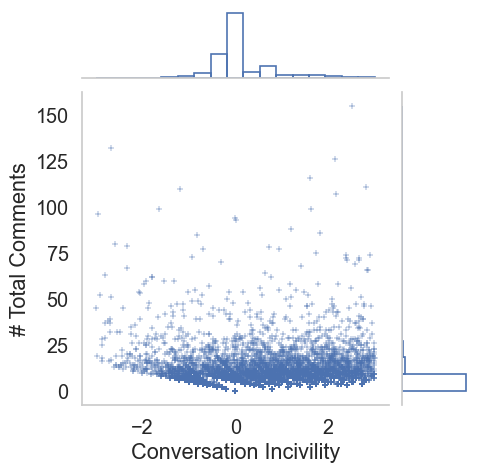}
			\caption{\# Total comments}
			\label{fig:total}
		\end{subfigure}
		\hfill
		\begin{subfigure}[b]{0.23\textwidth}
			\centering
			\includegraphics[width=\textwidth]{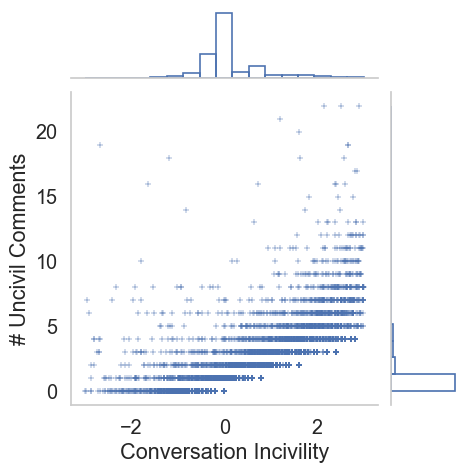}
			\caption{\# Uncivil comments}
			\label{fig:uncivil}
		\end{subfigure}
		\hfill
		\begin{subfigure}[b]{0.23\textwidth}
			\centering
			\includegraphics[width=\textwidth]{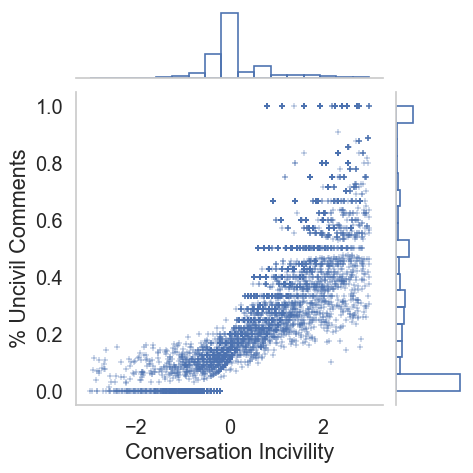}
			\caption{\% Uncivil comments}
			\label{fig:ratio}
		\end{subfigure}
		\caption{
			Comparison of our conversation incivility metric and three prior metrics:
			(a) number of total comments, (b) number of uncivil comments, and (c) percentage of uncivil comments.
			Our metric distinguishes between many conversations that obtain the same scores with prior metrics.
		}
		\label{fig:metric_compare}
	\end{figure}
	
	\begin{figure}[t]
		\centering
		\includegraphics[width=1\linewidth]{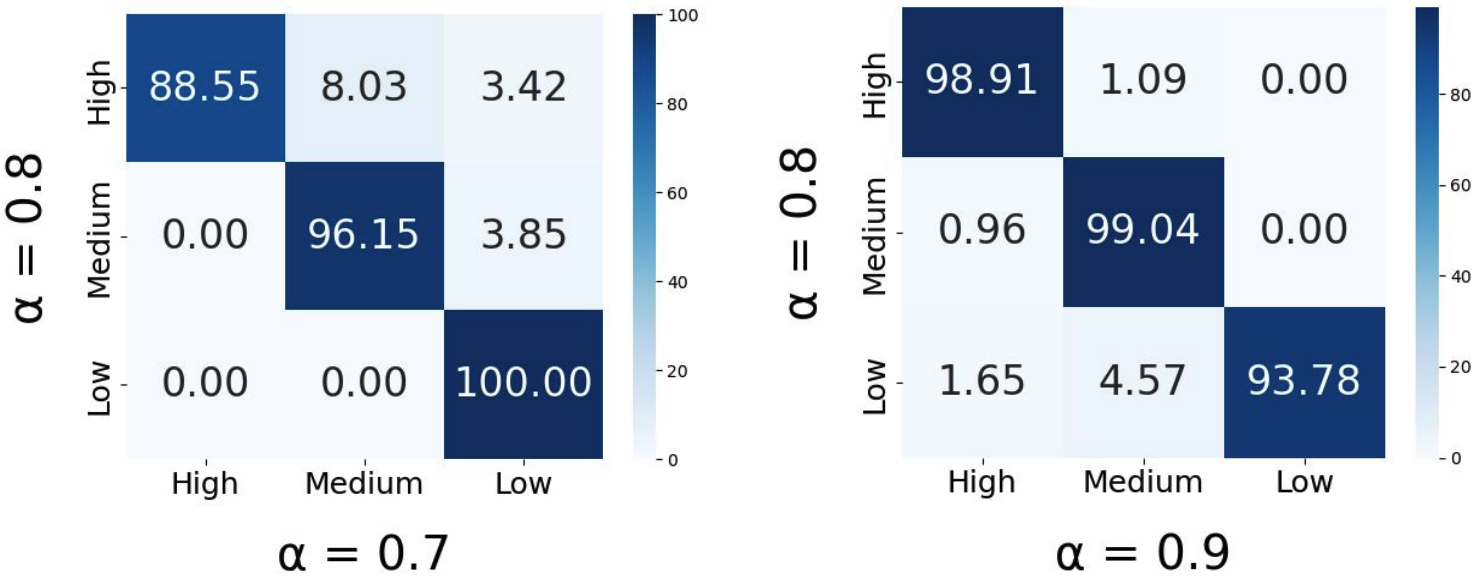}
		\caption{Confusion matrix (percentages) showing label changes (High, Medium, or Low incivility) when $\alpha$ is 0.8 vs. 0.7 and 0.9 respectively. } 
		\label{fig:alphas}
	\end{figure}
	
	\begin{table}
		\small
		\centering
		\begin{tabular}{l r}
			\toprule
			\multicolumn{2}{l}{Hate: \emph{Lol this thread is full of internet losers like you.}} \\
			\multicolumn{2}{l}{Reply: \emph{Ha! You don't even know me little man.}} \\ \addlinespace
			High incivility & $S(r) = 5.02$, [u:26, c:7] \\ 	
			\midrule
			\multicolumn{2}{l}{Hate: \emph{Too easy to trigger you maga f**ks, snowflake.}} \\
			\multicolumn{2}{l}{Reply: \emph{Have a nice day!}} \\ \addlinespace
			Medium incivility & $S(r) = 0$, [u:0, c:0] \\ 					
			\midrule
			\multicolumn{2}{p{7cm}}{Hate: \emph{Trash talking is perfectly ok. He is an a**hole.}} \\
			\multicolumn{2}{p{7cm}}{Reply: \emph{It's *tolerable* under those conditions. It is not perfectly okay.}}\\ \addlinespace
			Low incivility & $S(r) = -3.92$, [u:3, c:29]  \\ 
			\bottomrule
		\end{tabular}
		
		\caption{Examples from our corpus and their incivility levels.
			We also include their incivility scores and number of uncivil (u) and civil comments (c).}
		\label{t:corpus-examples}
	\end{table}
	
	The metric to measure conversation incivility following a reply $r$ requires us to get
	the number of uncivil and civil comments published after $r$ by each user.
	Compared with hateful comments, uncivil comments include broader cases~\cite{davidson-etal-2020-developing}.
	We therefore calculate uncivil comments based on the output of three classifiers.
	We build three models by training the same architecture as before with the corpus by \citet{qian-etal-2019-benchmark}
	and two additional corpora~\cite{hateoffensive,vidgen-etal-2021-introducing}.
	We consider a comment published after $r$ as uncivil if any of the three classifiers predicts \emph{uncivil}. 
	Otherwise, we consider it \emph{civil}.
	After calculating uncivil behavior and civil behavior after each reply,
	calculating the conversation incivility score $S(r)$ is straightforward.
	We experiment with $\alpha = 0.8$, 
	as we consider  one uncivil content more critical and several civil messages are needed to neutralize uncivil content.
	We do not claim that this is the best choice.
	Instead, we argue that $\alpha$ ought to be chosen based on the level of uncivil behaviors that is acceptable in a conversation, and can be adjusted according to the social media platform or tolerance to incivility. 
	As we shall see, $\alpha=0.8$ leads to almost perfect agreement with human annotators in our Reddit corpus.
	
	\newcolumntype{P}[1]{>{\centering\arraybackslash}p{#1}}
	\begin{table*}
		\centering
		\small
		\begin{tabular}{l P{.5in}P{.5in}P{.5in}P{.5in}P{.5in}P{.5in}}
			\toprule
			& All & Discussion & Hobby & Identity & Meme & Media\\ \midrule
			\textbf{Textual factors}\\
			~~~Tokens & $\uparrow\uparrow\uparrow$ & $\uparrow\uparrow\uparrow$ & $\underline\uparrow\underline\uparrow$ & $\uparrow\uparrow\uparrow$ & $\uparrow\uparrow\uparrow$ & $\uparrow\uparrow\uparrow$ \\ 
			~~~Negations & $\uparrow\uparrow\uparrow$ & $\uparrow\uparrow\uparrow$ & $\underline\uparrow$ &  $\uparrow\uparrow\uparrow$ &  $\uparrow\uparrow\uparrow$ & $\underline\uparrow\underline\uparrow\underline\uparrow$ \\
			~~~1st person pronouns & $\uparrow\uparrow\uparrow$ & $\underline\uparrow\underline\uparrow\underline\uparrow$ & $\underline\uparrow$ &  $\uparrow\uparrow\uparrow$ & $\uparrow\uparrow\uparrow$ & $\underline\uparrow\underline\uparrow$ \\
			~~~2nd person pronouns & $\uparrow\uparrow\uparrow$ & $\uparrow\uparrow\uparrow$ & $\uparrow\uparrow\uparrow$ & $\uparrow\uparrow\uparrow$ & $\uparrow\uparrow\uparrow$ & $\uparrow\uparrow\uparrow$ \\
			~~~Named entity (norp) & $\uparrow\uparrow\uparrow$ & $\uparrow\uparrow\uparrow$ & $\underline\uparrow$ & $\underline\uparrow$ & & $\underline\uparrow$ \\
			~~~Question marks & $\uparrow\uparrow\uparrow$ & $\underline\uparrow$ &  & $\underline\uparrow\underline\uparrow\underline\uparrow$ & $\underline\uparrow$ & $\underline\uparrow$ \\
			~~~Quotations & $\uparrow\uparrow\uparrow$  & $\uparrow\uparrow\uparrow$  & $\underline\uparrow$ & $\underline\uparrow\underline\uparrow$ &  & $\underline\uparrow$ \\
			\midrule
			\textbf{Sentiment factors}\\
			~~~Positive words & $\downarrow\downarrow\downarrow$ & $\downarrow\downarrow\downarrow$ & $\underline\downarrow\underline\downarrow\underline\downarrow$ & $\downarrow\downarrow\downarrow$ & $\underline\downarrow\underline\downarrow$ & $\underline\downarrow\underline\downarrow$ \\
			~~~Negative words & $\uparrow\uparrow\uparrow$ & $\uparrow\uparrow\uparrow$ & $\uparrow\uparrow\uparrow$ & $\uparrow\uparrow\uparrow$ & $\uparrow\uparrow\uparrow$ & $\uparrow\uparrow$ \\
			~~~Disgust words & $\uparrow\uparrow\uparrow$ & $\underline\uparrow\underline\uparrow\underline\uparrow$ & $\underline\uparrow\underline\uparrow\underline\uparrow$ & $\uparrow\uparrow\uparrow$  &  $\underline\uparrow$ & $\underline\uparrow$ \\
			~~~Hatred words & $\uparrow\uparrow\uparrow$ & $\underline\uparrow$ & & $\underline\uparrow\underline\uparrow$ & &  $\underline\uparrow\underline\uparrow\underline\uparrow$ \\
			~~~Angry words &  $\uparrow\uparrow\uparrow$ & $\underline\uparrow\underline\uparrow$ & $\underline\uparrow$ & $\underline\uparrow\underline\uparrow$ & $\underline\uparrow$ & $\underline\uparrow\underline\uparrow$ \\
			\bottomrule
		\end{tabular}
		
		\caption{Linguistic analysis comparing the replies to hateful comments that have high and low conversation incivility.
			We analyze all Reddit conversations (Column 2) and each subreddit category (Column 3-7, see the list in the Appendices).
			Number of arrows indicates the p-value (t-test; one: $p<0.05$, two: $p<0.01$, and three: $p<0.001$).
			Arrow direction indicates whether higher values correlate with high (up) or low (down) incivility. 
			Tests that \emph{do not pass} the Bonferroni correction are underlined.}
		\label{t:linguistic}
	\end{table*}

	The scatter plots in Figure~\ref{fig:metric_compare}
	compare our conversation incivility metric with three prior metrics:
	\begin{inparaenum}[(a)]
		\item number of (total) comments, 
		\item number of uncivil comments, and
		\item percentage of uncivil comments.
	\end{inparaenum}
	The three prior metrics often assign the same incivility scores to many conversations.
	For example, 
	the vast majority of conversations following a reply to hate speech are very short (few total comments, Figure \ref{fig:metric_compare}a)
	and do not have any uncivil comments (Figure \ref{fig:metric_compare}b).
	Similarly, the percentage of uncivil comments is low,
	although this metric is less biased towards low scores.
	Our conversation incivility metric (x-axis in the three plots)
	assigns different scores to conversations that receive the same scores with the three prior metrics,
	thereby providing a more nuance distinction of civil and uncivil conversations.


	
	\paragraph{Manual Validation}
	\label{ss:validation}
	To validate the conversation incivility scores obtained by our metric,
	we manually annotate a small benchmark.
	Specifically, we create a benchmark with ground-truth annotations in four steps.
	First, we randomly select 500 replies to hate comments from our corpus such that the 
	follow-up conversation has at least one comment.
	Second, we retrieve Reddit snippets containing
	the hateful comment,
	the reply,
	and the follow-up conversation.
	Third, we randomly pair the 500 snippets, resulting in 250 pairs.
	Fourth, annotators manually annotate which of the follow-up conversations in each pair is more uncivil.
	Two research assistants were hired as annotators.
	10 pairs were discarded due to uncertainty;
	annotators agreed on 194 of the remaining 240 pairs (80.8\%).
	The Cohen's $\kappa$ coefficient is 0.62, which is considered \emph{substantial} agreement~\cite{artstein2008inter}.
	We include in our benchmark the 194 pairs with perfect agreement.

	
	Armed with the benchmark,
	we compare the ground truth (i.e., which of the two follow-up conversations in each pair is more uncivil according to the human annotators)
	with (a)~our conversation incivility metric and (b) the best prior metric
	(i.e., percentage of uncivil comments, Figure \ref{fig:metric_compare}).
	Using the scores obtained with our metric (higher score indicates more uncivil),
	we match the ground truth in
	183 out of the 194 pairs (94.3\%), resulting in a 
	Cohen's $\kappa$ of 0.89, which is almost perfect agreement. 
	On the other hand,
	using the percentage of uncivil comments, we match the ground truth in
	172 out of the 194 pairs (88.7\%), resulting in a
	Cohen's $\kappa$ of 0.77, which is substantial agreement.
	Based on this evaluation, we conclude that while the percentage of uncivil comments is a valid choice to approximate conversation incivility,
	our conversation incivility metric more closely approximate humans judgments thus it is more sound.
	
	\paragraph{Conversation Incivility Level}
	We use the scores to group all replies in our corpus based on quantiles (top 25\%, middle 50\%, and bottom 25\%). 
	The score ranges are as follows:
	\begin{compactitem}
		\item \emph{High} incivility with $ S(r) \in (0.22,16.26] $;
		\item \emph{Medium} incivility  with $ S(r) \in (-0.20, 0.22] $; and
		\item \emph{Low} incivility with  $ S(r) \in [-48.76, -0.20]$.
	\end{compactitem}
	
	We refer to this grouping (high, medium or low) as the conversation incivility level.
	Figure~\ref{fig:alphas} shows the changes in incivility levels depending on other choices of $\alpha$ (0.7 and 0.9).
	Most conversations are assigned the same incivility level.
	We note that absolute incivility scores are not critical.
	Rather, incivility scores should be used as a means to compare the incivility of two conversations.
	The following analyses are based on the conversation incivility levels.
	
	\section{Corpus Analysis} 
	\label{s:corpus_analysis}
	Our final corpus consists of 34,115 replies to hateful comments from Reddit along with the conversation following each reply. 
	We show examples of each incivility level in Table \ref{t:corpus-examples}. 
	In the first example,
	the reply shows disagreement by attacking the author of the hate comment (\emph{little man}).
	This counter hate reply leads to high conversation incivility: 26 out of 33 comments that follow the reply are uncivil.
	Indeed, we observe the follow-up conversation is made up of behaviors where the original authors of the hateful comment and the reply repetitively denigrate each other.
	In the second example, the reply uses sarcasm to stop arguing with the hateful comment  (``\emph{Have a nice day!}'').
	There are no comments after this reply, yielding a 0 incivility score.
	It is followed by \emph{medium} conversation incivility---no additional comment is posted.
	In the third example, the reply denounces the misbehavior in the hate comment is inappropriate without getting into details or personal attacks (``\emph{[...] It is not perfectly okay.}'').
	This is a common counter hate strategy~\cite{DBLP:conf/icwsm/MathewSTRSMG019} and the follow-up conversation is barely uncivil: although 3 comments are uncivil, they are posted by the same user and 29 comments from 24 users remain civil.

	\paragraph{Linguistic insights} 
	We perform a linguistic analysis to shed light on differences in the language people use in the replies that elicit high and low conversation incivility (Table \ref{t:linguistic}).
	As different Reddit communities vary in content due to differences in topics, moderation rules, etc.~\cite{https://doi.org/10.48550/arxiv.2109.05152,Weld_Zhang_Althoff_2022}, 
	we further conduct analyses for each type of community to explore whether the differences between high and low incivility across communities exist.
	The 39 subreddits are grouped into five categories based on the taxonomy from~\citet{Weld_Zhang_Althoff_2022}.
	The authors hand-labeled communities in the categories iteratively until reaching agreement among them. 
	We include News in Media-sharing as both communities are for sharing information.
	The five categories are: Discussion (\emph{e.g., r/antiwork}), Hobby (\emph{e.g., r/dota2}), Identity (\emph{e.g., r/Feminism}), Meme (\emph{e.g., r/DankMemes}), and Media-sharing (\emph{e.g., r/worldnews}) (see Appendices).
	
	All factors we consider are based on counts of
	(a)~textual features (top block)
	or
	(b) presence of words related to sentiment.
	We consider a reply uses quotations if it has the character  `$>$' and the text that follows overlaps with the hateful comment~\cite{chakrabarty-etal-2019-ampersand,jo-etal-2020-detecting}.
	We check for negation cues using the list by~\citet{fancellu-etal-2016-neural}. 
	We use spaCy to recognize named entities.\footnote{\url{https://spacy.io/usage/linguistic-features}}
	For sentiment and cognition, we use the Sentiment Analysis and Cognition Engine (SEANCE) lexicon,
	a tool for psychological linguistic analysis \cite{crossley2017sentiment}. 
	Statistical tests are conducted using unpaired t-tests between two groups: the replies eliciting high or low incivility.
	We draw several interesting insights:
	\begin{compactitem}
		\item Regarding textual factors, we observe that among all the replies, the more tokens, negations, pronouns (1st and 2nd person), entities (nationalities, religious, or political groups), question marks and quotations, 
		the more likely the subsequent conversation of a reply to hateful comment is uncivil.
		Negation cues are often used to dispute the hateful comment.
		Presence of \emph{you} and \emph{your} usually refers to the author of the hateful comment.
		\item Regarding sentiment, there are significant differences in the uses of positive and negative words. 
		Replies that use more hatred, disgust and angry words tend to lead to more incivility in the follow-up conversations.
		\item Although topics and content vary across communities, we observe high consistency in the differences between high and low conversation incivility, especially for 2nd person pronouns and negative words.
	\end{compactitem}

	\section{Experiments and Results} 
	\label{s:experiments}
	
	We experiment with models to solve two problems:
	\begin{compactitem}
		\item Determining the conversation incivility level of a reply to hate speech: high, medium or low incivility; and
		\item Differentiating the top-$k$\% and bottom-$k$\% replies according to their conversation incivility scores.
	\end{compactitem}

	All our models are neural classifiers with the RoBERTa transformer~\cite{DBLP:journals/corr/abs-1907-11692} as the main component.
	We use the pretrained models by HuggingFace~\cite{wolf-etal-2020-transformers}
	and Pytorch~\cite{NEURIPS2019_9015} to implement our models.

	\subsection{Determining Incivility Level}
	\label{ss:levels}
	
	The neural architecture consists of the RoBERTa transformer, 
	a fully connected layer (768 neurons and \texttt{tanh} activation), 
	and 
	another fully connected layer (3 neurons and softmax activation) to make predictions (high, medium, or low incivility). 
	To investigate whether adding the hate comment would be beneficial, we consider three textual inputs:
	\begin{compactitem}
		\item the hate comment;
		\item the reply to the hate comment; and
		\item the hate comment and the reply.
	\end{compactitem}
	
	Intuitively, the reply is the most important input, but as we shall see including the hate comment is beneficial.
	We concatenate both inputs with the \texttt{[SEP]} special token. 
	
	\noindent
	\textbf{Pretraining with Related Tasks} 
	We experiment with several corpora to investigate whether pretraining with related tasks is beneficial. 
	Specifically, we pretrain with existing corpora annotating: 
	(a) hate speech: hateful or not hateful~\cite{hateoffensive}; 
	(b) sentiment: negative, neutral, or positive \cite{rosenthal-etal-2017-semeval}; 
	(c) sarcasm: sarcasm or not sarcasm \cite{ghosh-etal-2020-report}; 
	(d) counterspeech: hate, neutral, or counterhate \cite{yu-etal-2022-hate}; and
	(e) stance: agree, neutral, or attack \cite{pougubiyong2021debagreement}.

	\noindent
	\textbf{Blending Additional Annotations}
	Pretraining takes place prior to training with our corpus.
	We also experiment with a complementary approach: blending additional corpora during the training process, as proposed by \citet{shnarch-etal-2018-will}.
	With blending, there are two phases in the training process: 
	(a) $\mathit{m}$ blending epochs using all of our corpus and a fraction of an additional corpus,
	and
	(b) $\mathit{n}$ epochs using only our corpus. 
	In each blending epoch, a random fraction of an additional corpus is fed to the network.
	The fraction is determined by a blending factor $\alpha \in [0..1]$. 
	The first blending epoch is trained with our corpus and the whole additional corpus.
	Subsequent blending epochs use smaller fractions of the additional corpus.
	We use for blending purposes the corpora we use for pretraining that annotate three labels~\cite{rosenthal-etal-2017-semeval,pougubiyong2021debagreement,yu-etal-2022-hate}.
	
	\begin{table*}[t!]
		\small
		\centering	
		\begin{tabular}{l ccc ccc ccc ccc}
			\toprule
			\multicolumn{1}{c}{} & \multicolumn{3}{c}{High} & \multicolumn{3}{c}{Medium} & \multicolumn{3}{c}{Low} & \multicolumn{3}{c}{Weighted Average} \\
			\cmidrule(lr){2-4} \cmidrule(lr){5-7} \cmidrule(lr){8-10} \cmidrule(lr){11-13} 
			& P & R & F1 & P & R & F1 & P & R & F1 & P & R & F1 \\
			\hline
			\addlinespace[1pt]
			Majority Baseline & 0.00 & 0.00 & 0.00 & 0.49 & 1.00 & 0.66 & 0.00 & 0.00 & 0.00 &  0.24 & 0.49 & 0.32 \\ \addlinespace
			
			RoBERTa classifier with \\ 
			~~~hate comment & 0.34 & 0.31 & 0.33 & 0.53 & 0.72 & 0.61 & 0.27 & 0.10 & 0.15 & 0.41 & 0.46 & 0.42 \\ \addlinespace
			
			~~~reply & 0.42 & 0.33 & 0.37 & 0.53 & 0.77 & 0.63 & 0.29 & 0.10 & 0.15 & 0.44 & 0.49 & 0.44\\		
			~~~~~~+ blending & 0.42 & 0.42 & 0.42 & 0.55 & 0.76 & 0.63 & 0.33 & 0.08 & 0.13 & 0.46 & 0.50 & 0.45\\
			~~~~~~+ pretraining† & 0.44 & 0.37 & 0.40 & 0.56 & 0.72 & 0.63 & 0.32 & 0.20 & 0.25 & 0.47 & 0.50 & 0.47\\ 
			~~~~~~~~~+ blending†‡ & 0.47 & 0.39 & 0.43 & 0.57 & 0.70 & 0.63 & 0.34 & 0.25 & 0.29 & 0.49 & 0.51 & 0.49\\ 
			\addlinespace
			
			~~~hate comment + reply & 0.43 & 0.32 & 0.36 & 0.55 & 0.66 & 0.60 & 0.32 & 0.27 & 0.29 & 0.46 & 0.48 & 0.46\\
			~~~~~~+ blending† & 0.43 & 0.37 & 0.40 & 0.54 & 0.79 & 0.64 & 0.35 & 0.10 & 0.16 & 0.47 & 0.51 & 0.46\\
			~~~~~~+ pretraining†‡ & \textbf{0.52} & \textbf{0.43} & \textbf{0.47} & \textbf{0.59} & \textbf{0.77}& \textbf{0.67} & 0.42 & 0.23 & 0.30 & \textbf{0.53 }& \textbf{0.55} & \textbf{0.52}\\
			~~~~~~~~~+ blending†‡ & 0.48 & 0.42 & 0.45 & 0.59 & 0.72 & 0.65 & \textbf{0.38} & \textbf{0.27} & \textbf{0.32} & \textbf{0.51} & \textbf{0.53} & \textbf{0.52}\\
			\bottomrule
			
		\end{tabular}
		\caption{Results obtained with several models. 
			We indicate statistical significance (McNemar’s test \cite{mcnemar1947note} over the weighted average) as follows: 
			† indicates statistically significant ($p<0.05$) results with respect to the \emph{reply} model,
			and ‡ with respect to the \emph{hate comment + reply} model.
			Training with the \emph{hate comment + reply}
			coupled with pretraining with stance or both pretraining and blending stance yields the best results (F1: 0.52).}
		\label{t:model-results}
	\end{table*}

	\subsubsection{Quantitative Results}
	We split the 34,115 replies in our corpus into train (60\%), development (20\%) and test (20\%) splits.
	We present results with the testing split in Table \ref{t:model-results}.
	The majority baseline always predicts \emph{medium}.
	The remaining rows present results with different settings:
	using as input the \emph{hate comment}, the \emph{reply} or both without pretraining or blending, and also with
	pretraining, blending and both.
	We provide here results pretraining and blending with the most beneficial tasks:
	counterspeech (\emph{+ pretraining} when using \emph{reply}, and \emph{+ blending} using \emph{hate comment + reply}),
	and stance for the remaining ones.
	We tune the blending factor $\alpha$  with the training and development splits, like other hyperparameters.
	We found the optimal $\alpha$ to be 0.5 when only blending and 1.0 when pretraining and blending. 
	
	Using only the reply as input is a strong baseline: it substantially outperforms the majority baseline (F1: 0.44 vs. 0.32).
	Using both the hate comment and reply yields better results (F1: 0.46).
	Pretraining and blending yield better results compared with blending alone (\emph{reply}: 0.49 vs. 0.45, \emph{hate comment + reply}: 0.52 vs. 0.46).
	Also, pretraining or pretraining and blending are more beneficial when the input is both the hate comment and the reply.
	Finally, the networks (a) pretraining and (b) pretraining and blending using both hate comment and reply yield the best results (F1: 0.52).
	
	\begin{table*}[htp!]
		\small
		\centering	
		\begin{tabular}{lrl  cccc cccc ccc}
			\toprule
			\multicolumn{3}{c}{} & \multicolumn{3}{c}{Top-k\%} && \multicolumn{3}{c}{Bottom-k\%} && \multicolumn{3}{c}{Weighted Average}  \\ \cmidrule(lr){4-6} \cmidrule(lr){8-10} \cmidrule(lr){12-14} 
			& Size && P & R & F1 && P & R & F1 && P & R & F1  \\ \midrule
			$\mathit{k}=5$  &  3,234 &&0.77 & 0.74 & 0.76 && 0.72 & 0.75 & 0.74	&& 0.75 & 0.75 & 0.75 \\
			$\mathit{k}=10$ &  6,124 && 0.74 & 0.77 & 0.76 && 0.68 & 0.65 & 0.67 && 0.72 & 0.72 & 0.72 \\
			$\mathit{k}=15$ & 10,434 && 0.70 & 0.76 & 0.73 && 0.68 & 0.61 & 0.65 && 0.69 & 0.69 & 0.69 \\
			$\mathit{k}=20$ & 17,231 && 0.67 & 0.63 & 0.65 && 0.65 & 0.68 & 0.66  && 0.66 & 0.66 & 0.66 \\ \bottomrule
			
		\end{tabular}
		\caption{Experimental results differentiating the top-$k$\% and bottom-$k$\% replies to hateful comments according to the incivility scores in the follow-up conversations.
			We present results for several values of $k$.
			The results are higher than when also identifying replies that elicit \emph{medium} incivility.
			Additionally, it is easier to differentiate the replies with the highest and lowest incivility scores in the follow-up conversations: the smaller the $k$, the higher the weighted average.
		}
		\label{t:model-optimalk}
	\end{table*}

	\subsection{Differentiating between the Top-$k$\% and Bottom-$k$\% replies}
	\label{ss:topbottom}
	Although determining the incivility level of any replies to hateful comment is a worthwhile goal,
	differentiating between the top-$k$\% and bottom-$k$\% replies according to the incivility scores of their follow-up conversations
	may lead to better actionable knowledge.
	Indeed, replies eliciting the highest (top-$k$\%) or lowest (bottom-$k$\%) incivility scores are more informative than the large amount of replies that elicit conversations with in-between incivility scores.
	
	Table \ref{t:model-optimalk} presents the results with several $k$ values after 
	retraining one of the best performing systems from Table \ref{t:model-results} (\emph{hate comment + reply + pretraining}).
	Results show that the smaller the $k$, the easier it is to differentiate between the two kinds of replies.
	This is especially true when $k=5$ (F1: 0.75).
	The results are encouraging. 
	Indeed, replies eliciting the highest and lowest conversation incivility differ in language usage and the classifier can distinguish them. 
	Also, these replies are the ones with the potential to elicit the highest and lowest incivility in the subsequent conversation, thus are the most useful to identify.
	

	\begin{table*}
		\small
		\centering
		\begin{tabular}{@{\hspace{.03in}}p{3.0cm}p{0.1cm}p{8.6cm}ll@{\hspace{.03in}}}
			\toprule
			Error Type & \% & Example & Ground Truth  & Predicted  \\ \midrule
			
			Rhetorical question & 23 & Hate: \emph{You're living in the west. You're privileged.} &  & \\
			&    & Reply: \emph{Are you an idiot? Can you read? Feminist my a**. } & Medium & High  \\
			\midrule
			Uncivil reply followed by conversation with & 18 & Hate: \emph{I've addressed this about forty times with as many smooth brains as you so. You're an idiot.} & & \\
			low incivility	&   &  Reply: \emph{I think you might be the idiot here retard.} & Low & High \\
			\midrule
			Civil reply followed by conversation with  & 16 & Hate: \emph{You’re an ignorant twat who just parrots what they read in FB and reddit memes. [\ldots] What a cancer you are.}  & & \\
			high	incivility&  & Reply: \emph{Calling others cancer is taking it too far. Mind rule 4, please.} & High & Low\\
			\midrule
			Sarcasm or irony & 15 & Hate: \emph{No you retard, where is the f**king lie?} & & \\
			&  & Reply: \emph{Name calling nice argument.} & High & Low \\
			\midrule
			General knowledge & 10 & Hate: \emph{lol bet you thought a single thing you said wasn’t retarded.} & & \\
			&  & Reply: \emph{This place is infested with incels and TD trolls. } & High & Medium \\
			\midrule
			Negation & 8 & Hate: \emph{Why we have to tolerate Islam? They call us filth. Christians are horrible as well. Both are f**king awful.} && \\
			& & Reply: \emph{Not all Muslims are bigots, like not all Christians are bigots.} & Medium & High \\
			\bottomrule
		\end{tabular}
		\caption{Most common error types made by the best model (predictions by \emph{hate comment + reply + pretraining}).
		}
		\label{t:error}
	\end{table*}

	\section{Qualitative analysis}
	\label{s:erroranalysis}
	
	When determining the incivility level of a reply,
	when does our best model (Table \ref{t:model-results}) make mistakes? 
	To investigate this question,
	we manually analyze 200 random samples in which the output of the network differs from the ground truth. 
	Table \ref{t:error} exemplifies the most common error types.
	
	The most frequent error type (23\%) is \emph{Rhetorical questions},
	a finding consistent with previous work~\cite{schmidt-wiegand-2017-survey}. 
	In the example,
	the model fails to realize that the question in the reply is used to point out inappropriate content rather than expecting an answer.
	There are no comments after the reply, therefore the conversation that follows has medium incivility.
	
	The second and third most common error types (18\% and 16\%)
	occupied when a reply is
	(a)~uncivil but followed by a conversation with low incivility
	or
	(b)~civil but followed by a conversation with high incivility.
	Using uncivil language is a counter hate strategy~\cite{DBLP:conf/icwsm/MathewSTRSMG019},
	and the model fails to recognize when doing so leads to low conversation incivility.
	Similarly, the model struggles when countering hate politely elicits additional uncivil behaviors.
	When correcting misstatements, language toxicity may increase~\cite{10.1145/3411764.3445642}. 
	
	Sarcasm and irony are also common error types (15\%) in our task.
	This finding is consistent with previous work on 
	detecting hate~\cite{nobata2016abusive,qian-etal-2019-benchmark}.
	In the example, using sarcasm to point out a bad argument (i.e., name calling) elicits further hate and the model errs.
	
	Errors may also occur (10\%) when general knowledge is required to identify hate content that does not use offensive language (e.g., calling people \emph{incels}).
	
	Finally, we observe that \emph{negation} appears in 8\% errors.
	In the example, negations are used to point out the flaws of generalizing.
	We hypothesize that the model fails to identify that the reply is followed by medium incivility: negation does indicate high incivility in general (Table~\ref{t:linguistic}).

	\section{Conclusion and Discussion}
	In this work, we present a metric to assess conversation incivility and apply it to conversations following replies to hate posts in a large Reddit dataset. 
	Our metric takes into account the number of both civil and uncivil comments as well as the unique authors.
	A manual validation shows that out metric approximates human judgments better than previous proposals.
	Regardless of whether replies convincingly counter a hate post,
	we believe it is worthwhile to identify what kind of user-generated content attracts attention and shapes civil discussions.
	While we make no causal claims about which linguistic features could affect conversation incivility, 
	we show that the language of user-generated replies differs depending on their conversation incivility levels.
	The outcomes of the linguistic analysis are intuitive.
	For example, replies that use more negative and disgust words result in follow-up conversations with higher incivility.
	This insight is consistent across several communities.
	Experimental results show that pretraining with and blending existing corpora yield improvements,
	yet the task of forecasting conversation incivility is still challenging to automate.
	
	Experimental results with classifiers built to distinguish the
	replies that elicit follow-up conversation with the top-$k$\% and bottom-$k$\% incivility scores are encouraging.
	The smaller the $k$, the easier the task, despite the classifiers do not have access to the follow-up conversation.
	The work presented here opens the door to automated methods
	to forecast the incivility of the conversation following a reply to hate content---\emph{at the time} the reply is posted.
	It also may inform the design of effective strategies to mitigate the spread of hate \emph{without} having to censor hateful content.
	
	
	\section{Broader Perspectives, Ethics, and Competitive Interests}
	Our work provides both theoretical and empirical guidelines to assist online media in measuring, understanding, identifying, and even intervening with content that could elicit additional incivility.
	First, our research proposes a more comprehensive approach to measure conversational outcomes with regard to incivility, compared with traditional approaches using only the number or ratio of uncivil comments. 
	Second, our study provides language characteristics that could trigger additional incivility, revealing deeper insights into the nature of these conversations. 
	
	This work may motivate the design of systems that highlight content likely to elicit uncivil behaviors. 
	Doing so could assist moderation. 
	For instance, content that does not contain swear words or use implicit hate speech~\cite{elsherief-etal-2021-latent} have instigated additional uncivil behaviors and violated Reddit content policy.\footnote{\url{https://www.redditinc.com/policies/content-policy}}
	This poses a challenge for most detection systems. 
	However, our method could help identify this kind of content.
	In addition, our metric to estimate conversation incivility offers a new way to measure conversational outcomes, focused on the future health of the conversation.
	Our metric may be applied to broader topics (e.g., persuasion, online debate) and other online discussion forums.
	Finally, our work may also help design guidelines for how to appropriately engage in online discussions
	(e.g., pointing out what kinds of strategies are the most effective when countering online hatred~\cite{DBLP:conf/icwsm/MathewSTRSMG019}).

	\paragraph{Limitations}
	This work is not without limitations.
	First, our linguistic analyses should not be interpreted as  causal statements.
	Small-scale user studies could provide an understanding of how humans perceive hate speech and replies to hate speech.
	Additionally, some of the linguistic analyses relies on automated tools (e.g., spaCy)
	that do not always output the correct predictions (e.g., some NORP named entities are false positives).
	Second, we identify uncivil comments automatically with classifiers.
	These classifiers obtain good results but are not perfect.
	Third, we focus on forecasting conversation  ity only from language (the hate content and the reply).
	A promising line of future research could consider
	combining structural and linguistic features and incorporate other factors such as
	roles and user traits.
	Fourth, we only work with Reddit posts. Our research agenda also includes investigating generalizations to other online platforms.
	Finally, we only consider the hateful comment and the reply in our experiments.
	More complex modeling that takes into account additional context (e.g., the full conversation prior to the reply) may be beneficial.
	
	\paragraph{Ethical Considerations}
	\label{sec:ethical}
	The study has been through careful consideration of the risks and benefits to ensure that it is conducted in an ethical manner.
	First, we use the PushShift API to collect data from Reddit.\footnote{\url{https://pushshift.io/api-parameters/}} 
	The collection process is consistent with Reddit's Terms of Service. 
	Second, in contrast to private messaging services, Reddit is considered a public space for discussion~\cite{vidgen-etal-2021-introducing}. 
	It does not require IRB review.
	Users consent to have their data made available to third parties including academics when they sign up. 
	Ethical guidelines state that in this situation explicit consent is not required from each user \cite{DBLP:conf/tto/ProcterWBHEWJ19}. 
	We obfuscate user names to reduce the possibility of identifying users. 
	In compliance with Reddit's policy, we would like to make sure that our dataset will be reused for non-commercial research only.\footnote{\url{https://www.reddit.com/wiki/api-terms/}}
	Third, the annotators were warned of the potential hateful content before working on our task. 
	They were also encouraged to stop the annotation process whenever feel upset.
	We provide annotators with access to supporting services throughout the task.
	Annotators were compensated with \$8 per hour.
	Finally, we acknowledge the risk associated with releasing the dataset.
	However, we believe the benefit of shedding light on what replies elicit additional incivility outweighs any risks associated with the dataset release.

	\bibliography{aaai24}
	\appendix

	\section{Subreddit List}
	We provide here the list of subreddits we work with and the categories they belong to:
	\begin{itemize}
		\item Discussion: \textit{r/antiwork}, \textit{r/changemyview}, \textit{r/NoFap}, \textit{r/Seduction}, \textit{r/PurplePillDebate}, \textit{r/ShitPoliticsSays}, \textit{r/PurplePillDebate}, \textit{r/bindingofisaac}, \textit{r/FemaleDatingStrategy}, \textit{r/SubredditDrama};
		\item Hobby:  \textit{r/KotakuInAction}, \textit{r/DotA2}, \textit{r/technology}, \textit{r/modernwarfare},  \textit{r/playrust}, \textit{r/oblivion};
		\item Identity: \textit{r/bakchodi}, \textit{r/Feminism}, \textit{r/PussyPass}, \textit{r/MensRights}, \textit{r/Sino}, \textit{r/BlackPeopleTwitter}, \textit{r/india}, \textit{r/PussyPassDenied}, \textit{r/TwoXChromosomes}, \textit{r/GenZedong}, \textit{ r/antheism};
		\item Meme: \textit{r/4Chan}, \textit{r/justneckbeardthings}, \textit{r/HermanCainAward}, \textit{r/MetaCanada}, \textit{r/DankMemes}, \textit{r/ShitRedditSays};
		\item Media: \textit{r/conspiracy},  \textit{r/worldnews}, \textit{r/Drama}, \textit{r/TumblrInAction},  \textit{r/lmGoingToHellForThis}, \textit{r/TrueReddit}.
	\end{itemize}

	\section{Choice of $f$ Function}
	\begin{figure}
		\centering
		\begin{subfigure}[b]{0.23\textwidth}
			\centering
			\includegraphics[width=\textwidth]{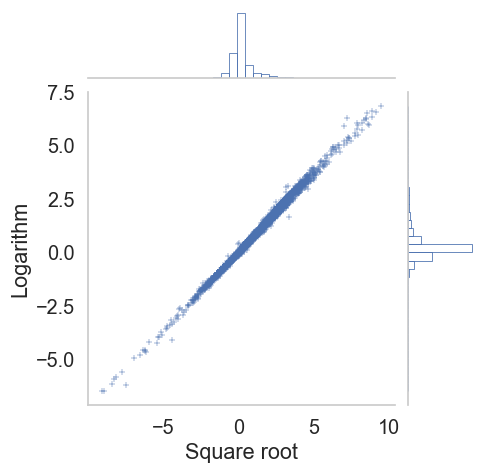}
			\caption{Logarithm}
			\label{fig:function}
		\end{subfigure}
		\hfill
		\begin{subfigure}[b]{0.23\textwidth}
			\centering
			\includegraphics[width=\textwidth]{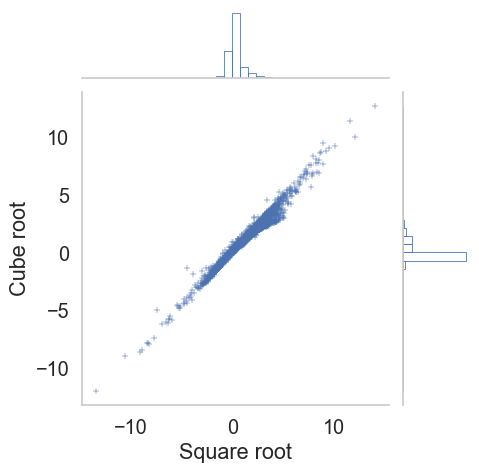}
			\caption{$\sqrt[3] x$}
			\label{fig:cube}
		\end{subfigure}
		\hfill
		\begin{subfigure}[b]{0.23\textwidth}
			\centering
			\includegraphics[width=\textwidth]{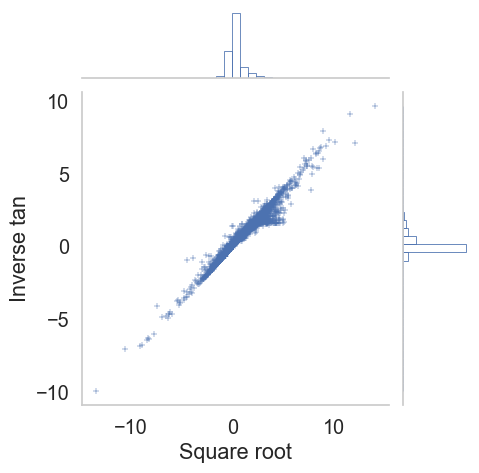}
			\caption{$\arctan x$}
			\label{fig:tan}
		\end{subfigure}
		\begin{subfigure}[b]{0.23\textwidth}
			\centering
			\includegraphics[width=\textwidth]{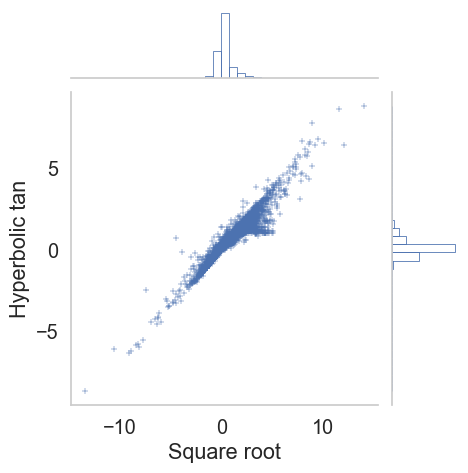}
			\caption{$\tanh x$}
			\label{fig:hyper}
		\end{subfigure}
		\caption{Comparison of incivility scores obtained with our choice of $f$ function ($f(x) = \sqrt{x}$
			and other strictly increasing concave down functions passing through the origin:
			logarithm ($\log (x+1)$),
			cube root ($\sqrt[3]{x}$),
			inverse tangent ($\arctan(x)$),
			and hyperbolic tangent ($\tanh(x)$).	
			Spearman's rank correlation coefficients are greater than 0.95 ($p<0.001$),
			indicating the the choice of $f$ is not critical, especially since incivility scores are meant to be used for comparison purposes (i.e., relative terms).}
		\label{fig:function_compare}
	\end{figure}
	
	\begin{table*}[htp!]
		\small
		\centering
		
		\begin{tabular}{l ccc ccc ccc ccc}
			\toprule
			\multicolumn{1}{c}{} & \multicolumn{3}{c}{High} & \multicolumn{3}{c}{Medium} & \multicolumn{3}{c}{Low} & \multicolumn{3}{c}{Weighted Average} \\
			\cmidrule(lr){2-4} \cmidrule(lr){5-7} \cmidrule(lr){8-10} \cmidrule(lr){11-13} 
			& P & R & F1 & P & R & F1 & P & R & F1 & P & R & F1 \\
			\hline
			Majority Baseline & 0.00 & 0.00 & 0.00 & 0.49 & 1.00 & 0.66 & 0.00 & 0.00 & 0.00 &  0.24 & 0.49 & 0.32 \\ 
			Trained with reply \\
			~~~~~~~~+ pretraining + blending\\
			~~~~~~~~~~~~Mean &0.48&	0.40&	0.43&	0.56&	0.71&	0.62&	0.33&	0.23&	0.28&	0.49&	0.49&	0.48\\
			~~~~~~~~~~~~(SD) & 0.03	& 0.03&	0.01&	0.01&	0.03&	0.01&	0.01&	0.04&	0.01&	0.01&	0.02&	0.01\\
			Trained with hate comment + reply\\
			~~~~~~~~+ pretraining + blending\\
			~~~~~~~~~~~~Mean&0.47&	0.38&	0.44&	0.60&	0.74&	0.67&	0.36&	0.25&	0.30&	0.52&	0.52&	0.52\\
			~~~~~~~~~~~~(SD) & 0.02&	0.05&	0.01&	0.01&	0.03&	0.02&	0.02&	0.05&	0.02&	0.01&	0.01&	0.01 \\
			\bottomrule
			
		\end{tabular}
		\caption{Detailed results (P, R, and F) predicting whether the \emph{reply} is folllowed by a conversation with High or Medium or Low incivility level when the input is only the reply or the hate comment. The results are using both pretraining and blending on related tasks. We experiment with multiple runs using different random seeds and report the mean scores and their standard deviation. }
		\label{t:multipleruns-results}
	\end{table*}
	
	Our conversation incivility metric uses a strictly increasing concave down $f$ function passing through the origin.
	We work with $f(x) = \sqrt{x}$,
	but choice of $f$ is not critical.
	Figure \ref{fig:function_compare} compares the incivility scores obtained with $f(x) = \sqrt{x}$ and four alternative $f$ functions:
	logarithm ($\log (x+1)$),
	cube root ($\sqrt[3]{x}$),
	inverse tangent ($\arctan(x)$),
	and hyperbolic tangent ($\tanh(x)$).	
	Spearman's correlation coefficients are 0.99, 0.99, 0.99, and 0.97 respectively ($p<0.001$).
	Thus any function would yield the same outcomes if incivility scores, as we recommend, are to be used only for comparative purposes (i.e., in relative rather than absolute terms).
	
	\section{Detailed Results}
	Table \ref{t:multipleruns-results} presents the mean scores of Precision, Recall and weighted F1-score and their standard deviation when we use both  pretraining and blending on related tasks with different random seeds. The results are consistent with the findings in our study: adding the \textit{hate comment} does improve the performance compared with the system that does not (0.52 vs. 0.48).
	Experiments were conducted on a Ubuntu server with two NVDIA TU102 GPUs using PyTorch 1.10.

\end{document}